\documentclass[aps,prl,showpacs,twocolumn]{revtex4}

\usepackage{graphicx}

\def\bra{\langle}
\def\ket{\rangle}

\def\cF{{\cal F}}

\def\cN{{\cal N}}

\def\vr{{\bf r}}

\def\vz{{\bf z}}

\def\gsw{\psi_{\rm gs}}
\def\GPw{\psi_{\rm GP}}
\def\td{{\rm 2D}}

\begin{document}

\title{Structure of excited vortices with higher angular momentum in Bose-Einstein condensates}

\author{Jian-Ming Tang}

\affiliation{Department of Physics and Astronomy, University of Iowa,
  Iowa City, IA 52242-1479}

%\date{\today}

\begin{abstract}
  
  The structure of vortices in Bose-Einstein condensed atomic gases is
  studied taking into account many-body correlation effects. It is
  shown that for excited vortices the particle density in the vortex
  core increases as the angular momentum of the system increases. The
  core density can increase by several times with only a few percent
  change in the angular momentum. This result provides an explanation
  for the observations in which the measured angular momentum is
  higher than the estimation based on counting the number of vortices,
  and the visibility of the vortex cores is simultaneously reduced.
  The calculated density profiles for the excited vortices are in good
  agreement with experimental measurements.
 
\end{abstract}

\pacs{67.40.Vs, 03.75.Fi, 02.30.Rz}

\maketitle

When a superfluid is put into rotation, vorticity is split into
discrete vortex lines rather than continuously distributed as in the
case of solid-body rotation~\cite{Onsager1949,Feynman1955}. The
dynamics, such as formation, reconnection and decay, of vortex lines
is influenced by the microscopic structure of the vortex
core~\cite{Donnelly1991}. Direct imaging of rotating Bose-Einstein
condensed atomic gases (BECs) has revealed, for the first time, the
particle density profile of the vortex core in neutral
superfluids~\cite{Matthews1999b,Madison2000a,Abo-Shaeer2001,Hodby2001a,Engels2002}.
The particle density is reduced in the core, and the density dips in
the measured images are used to identify the presence of vortex lines
in the experiments. In the mean-field approximation, known as the
Gross-Pitaevskii (GP) theory~\cite{Gross1961,Pitaevskii1961}, the
predicted density at the center of the vortex core is exactly zero
because the phase of the single-particle wave function at the vortex
line is not well-defined.  However, the observed density at the core
center is finite and varies in a wide range with different
experimental conditions. In many cases the core densities are quite
large, and cannot simply be accounted for by the finite resolution of
the imaging systems or by thermal excitations. Theoretically, finite
densities at the core centers arise from quantum fluctuations of the
locations of vortex lines due to many-body
correlations~\cite{Fetter1971a,Fetter1972,Tang2000,Tang2001b}.  The
fact that the core density is large and varies from measurements to
measurements further suggests that these vortex lines may be in
different vibrating modes~\cite{Kelvin1880} rather than straight. The
estimated density at the core center for a straight vortex line after
taking into account quantum fluctuations is comparable to the density
of particles depleted from the condensate in the ground state of a
BEC~\cite{Fetter1971a,Fetter1972}, which is not sufficient to explain
the observed large core densities.  In this Letter, the structure of
excited vortices is studied using many-body wave functions that
incorporate quantized motions of the vortex lines.  It is proposed
that the observed vortices with large core densities are in vibrating
modes corresponding to excited rotational states of BECs with an
angular momentum higher than the stationary GP states.  This proposal
is based on the results that the core density increases rapidly as the
angular momentum increases, and is supported by the experimental
findings that the angular momentum per particle, averaged over an
ensemble of one-vortex systems, is larger than
$\hbar$~\cite{Chevy2000a,Hodby2003}.  The existence of vortices with a
higher angular momentum also provides an explanation for the
observations in which the measured angular momentum is higher than the
estimation based on counting the number of vortices, and the
visibility of the vortex cores is simultaneously
reduced~\cite{Bretin2003}. The calculated density profiles for the
excited vortices are in good agreement with experimental measurements.

\begin{table}

\caption{
Ratios between the interatomic spacing, $\sigma$, and the healing length,
$\xi$, at the center of the traps in different experiments.
In the Thomas-Fermi approximation, the ratio is given by $\sigma/
\xi=2\left(15\pi^5Nm^3a_s^6\omega_\perp^2\omega_\parallel/ \hbar^3\right)^{1/15}$, where $N$ is the
total number of particles, $m$ is the mass, $a_s$ is the $s$-wave
scattering length, and $\omega_\perp$ and $\omega_\parallel$ are the trapping frequencies in
the perpendicular and parallel directions to the rotation axis.  The
last column shows the ratio for liquid helium assuming that the vortex
core radius is $1$ \AA.
}
\label{tb:ratio}

\centering
\begin{tabular}{|c|c|c|c|c|c|c|c|}
\hline
Atoms & $^{87}$Rb\cite{Matthews1999b} & $^{87}$Rb\cite{Madison2000a} & $^{23}$Na\cite{Abo-Shaeer2001} & $^{87}$Rb\cite{Hodby2001a} & $^{87}$Rb\cite{Engels2002} & $^4$He \\
\hline
$\omega_\perp/2\pi$ (Hz) & $7.8$ & $219$ & $84$ & $62$ & $8.35$ & - \\
\hline
$\omega_\parallel/2\pi$ (Hz) & $7.8$ & $11.7$ & $20$ & $175$ & $5.45$ & - \\
\hline
$N$ & $3\times 10^5$ & $10^5$ & $5\times 10^7$ & $2\times 10^4$ & $10^6$ & - \\
\hline
$\sigma/ \xi$ & $0.6$ & $0.9$ & $0.9$ & $0.8$ & $0.6$ & 3.5 \\
\hline
\end{tabular}

\end{table}

The importance of many-body correlations in the vortex core can be
seen in the following way.  In the GP theory, the description of a
vortex line is ``classical'' in the sense that the position and
velocity of a vortex line can be simultaneously determined. A quantum
description would be required if the position uncertainty of a vortex
line is comparable to the size of the vortex core. The position
uncertainty of a vortex line is given by the interatomic
spacing~\cite{Tang2001a,Tang2001b}, because the total angular momentum
of the system would be altered by one $\hbar$ if a vortex line fluctuates
across an atom in the superfluid. The size of the vortex core is given
by the healing length which is determined through a balance between
the kinetic and potential energies for a density gradient.  Although
the healing length can be much larger than the interatomic spacing in
the weakly interacting limit, they are often comparable with each
other in realistic situations (See Table~\ref{tb:ratio} for
comparisons). As a result, the quantum fluctuations of vortex lines
are generally important.  We should note that it is possible to
address the finite column densities within the framework of the GP
theory by considering bent vortex lines~\cite{Garcia-Ripoll2001a}.
However, such considerations are limited to metastable states with
broken rotational symmetry and with a lower angular momentum.

For simplicity I consider a rotating BEC with only a single vortex
point in two dimensions, but the formulation can be applied to a
vortex line in a three-dimensional system. To obtain a quantum
description for the vortex, many-body wave functions are constructed
as linear combinations of the GP wave functions parameterized by the
location of the vortex~\cite{Tang2001a,Tang2001b}.  The GP wave
functions are used as basis states in analogous to the position space
representation of a particle in quantum mechanics. For a $N$-particle
system, such a wave function is written as,
\begin{eqnarray}
\Psi(\vr_1,\cdots\vr_N) & = & \int d^2\vr_0F(\vr_0)\prod_{i=1}^N\GPw(\vr_i;\vr_0) \;,
\label{eq:MBW}
\end{eqnarray}
where $F(\vr_0)$ is the weight function that represents the effective
dynamics of the vortex, and $\GPw(\vr;\vr_0)$ is the normalized
solution of the time-independent GP equation with a vortex located at
$\vr_0=(r_0,\theta_0)$. When Eq.~(\ref{eq:MBW}) is generalized to describe
a vortex line in three dimensions, the weight function becomes an
effective ``wave function'' of a string rather than a particle.  For
the case of a straight vortex line with the zero-point motion, the
wave function in Eq.~(\ref{eq:MBW}) is similar to the shadow wave
functions used to study vortices in liquid $^4$He~\cite{Sadd1997}.

The weight functions for the energy eigenstates of the vortex can be
solved by diagonalizing the many-body Hamiltonian within the sub
Hilbert space spanned by the GP basis states. In the case that the
system has a uniform ground state with a density $\rho_{\rm 2D}=\sigma^{-2}$, the
weight functions are found~\cite{Tang2001b} to be
\begin{eqnarray}
F_{n,l}(\vr_0) & = & \cN_{n,l}\,x^{|l|}e^{-x^2/2}e^{il\theta_0}\,_1\cF_1(a,b;x^2) \;, \label{eq:Usol}
\end{eqnarray}
where $n\geq 0$, $n\geq l\geq -N$, $x=\sqrt{\pi}r_0/\sigma$, $\cN_{n,l}$ is the
normalization constant determined by $\bra\Psi_{n,l}|\Psi_{n,l}\ket=1$,
$a=-(2n-l-|l|)/2$, $b=|l|+1$ and $_1\cF_1(a,b;x^2)$ is the confluent
hypergeometric function~\cite{Abramowitz1977}. It is straightforward
to verify that the state $|\Psi_{n,l}\ket$ is the eigenstate of the
angular momentum with eigenvalue $(N+l)\hbar$, given that the phase of
$\GPw(\vr;\vr_0)$ takes the form of $\theta+f(\theta-\theta_0)$ in azimuthally
symmetric systems. The weight function $F_{0,0}(\vr_0)$ is a Gaussian
centered at the origin, and the state $|\Psi_{0,0}\ket$ is the
corresponding many-body state that includes the zero-point motion to
the Hartree state of a centered vortex. The weight functions in
Eq.~(\ref{eq:Usol}) have exactly the same form as the energy
eigenfunctions of a charged particle in a constant magnetic
field~\cite{Landau1977}.  This is because the motion of a vortex is
also driven by a velocity-dependent transverse force like the Lorentz
force.  For nonuniform systems such as BECs in harmonic traps, the
exact solutions for the weight functions are not known.  However, in
the case that the radius of the system is much larger than the size of
the vortex core and the vortex is near the trap center, the weight
functions can be approximated by Eq.~(\ref{eq:Usol}) with $\rho_\td$
equals to the local density of the ground state~\cite{Tang2001b}.
Therefore, I will use the wave function in Eq.~(\ref{eq:MBW}) with
the weight functions given by Eq.~(\ref{eq:Usol}) to study the
structure of vortex cores in trapped BECs.

To use Eq.~(\ref{eq:MBW}), the first step is to set up the GP basis
states.  In order to solve the GP equation for a system with a given
number of particles, it is convenient to choose the length unit to be
the healing length, $\xi=\sqrt{\sigma^3/8\pi a_s}$, where $\sigma$ is the
interatomic spacing of the ground state at the trap center, and $a_s$
is the s-wave scattering length characterizing the interaction
strength between particles. The ground state wave function
$\bar\gsw(r)$ is first solved with the boundary condition
$\bar\gsw(0)=1$. This particular choice of boundary condition and the
length unit have the advantage that the normalization of the wave
function is related to the total number of particles as
\begin{eqnarray}
N & = & \frac{1}{\sigma^2}\int d^2\vr\left|\bar\gsw(r)\right|^2 \;.\label{eq:Np2D}
\end{eqnarray}
Then the GP state that has a centered vortex and the same number of
particles is solved and written as
\begin{eqnarray}
\bar\GPw(\vr;{\bf 0}) & = & g(r)\bar\gsw(r)e^{i\theta} \;,
\end{eqnarray}
where $g(r)$ is the amplitude ratio of the vortex state to the ground
state. The normalized GP basis states with an off-centered vortex can
now be approximated by
\begin{eqnarray}
\GPw(\vr;\vr_0) & \approx & \frac{1}{\sqrt{N}\sigma}g(|\vr-\vr_0|)\bar\gsw(r)e^{i\phi(\vr;\vr_0)} \;,
\end{eqnarray}
where the phase satisfies the following set of linear differential
equations,
\begin{eqnarray}
  \nabla\times\nabla\phi(\vr;\vr_0) & = & 2\pi\,\delta(\vr-\vr_0) \;,\label{eq:PDE1} \\
  \nabla\cdot\left[|\GPw(\vr;\vr_0)|^2\nabla\phi(\vr;\vr_0)\right] & = & 0 \;.\label{eq:PDE2}
\end{eqnarray}
The first equation specifies that the vortex is located at $\vr_0$,
and the second is the continuity equation which directly follows from
the GP equation. The general solutions of Eq.~(\ref{eq:PDE1}) can be
written as $\phi(\vr;\vr_0) = \vartheta(\vr-\vr_0)+\tilde\phi(\vr;\vr_0)$, where
$\vartheta(\vr-\vr_0)$ is the azimuthal angle of the vector $\vr-\vr_0$, and
$\tilde\phi(\vr;\vr_0)$ is a single-valued function that remains to be
determined by Eq.~(\ref{eq:PDE2}).

\begin{figure}

\includegraphics[width=\columnwidth]{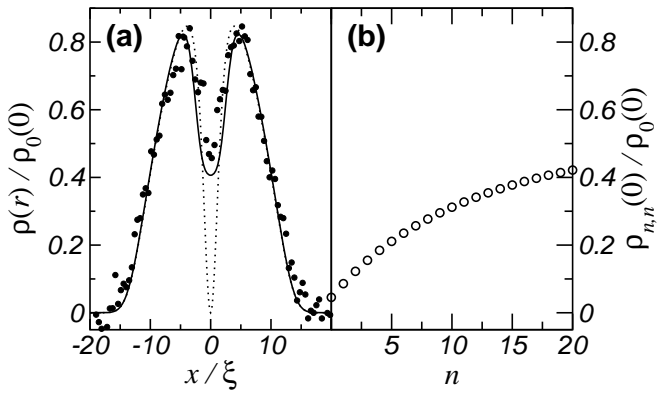}
\caption{(a) The integrated column density profile of a rotating BEC
  with a vortex at the center. The solid dots show the experimental
  data in Ref.~\cite{Madison2000a}. The solid line shows the density
  profile of the excited state ($n=18$) whose angular momentum per
  particle is about $1.05\hbar$. I approximate the 3D column density
  using the 2D results as $\rho_{\rm col}(r)=\rho_{\rm 2D}(r)\bar\gsw(r)$.
  The density profile of the GP state is shown as the dotted line for
  comparison. $\rho_0(0)$ is the central density of the ground state. (b)
  The open circles show the scaling behavior of the central density
  with respect to the angular momentum. }
\label{fig:denVIH}

\end{figure}

In the case that the vortex is close to the trap center, the
correction term $\tilde\phi(\vr;\vr_0)$ is only significant in the
region, $r>r_0$. In this region, the general solution of
Eq.~(\ref{eq:PDE1}) can be expanded as
\begin{eqnarray}
  \phi(\vr;\vr_0) & = & \theta+\sum_{j=1}^\infty \left(\frac{r_0}{r}\right)^j\left[\tilde\phi_j(r)+\frac{1}{j}\right]\sin[j(\theta-\theta_0)] \;,\nonumber\\ \label{eq:GS1}
\end{eqnarray}
where the factor $1/j$ comes from the azimuthal angle $\vartheta(\vr-\vr_0)$.
Since the density profile of the vortex state in this region is not
far from the ground state density profile, we can replace
$|\GPw(\vr;\vr_0)|^2$ in Eq.~(\ref{eq:PDE2}) with the density profile of the
ground state, $\rho_0(r)=|\gsw(r)|^2$. By substituting Eq.~(\ref{eq:GS1})
into Eq.~(\ref{eq:PDE2}), the differential equation for each
$\tilde\phi_j(r)$ is obtained as follows,
\begin{eqnarray}
\!\!\!\!\!\left[\frac{1}{r}\frac{d}{dr}r\frac{d}{dr}+\frac{\rho_0^\prime(r)}{\rho_0(r)}\frac{d}{dr}-\frac{j^2}{r^2}\right]\frac{\tilde\phi_j(r)}{r^j} & = & \frac{1}{r^{j+1}}\frac{\rho_0^\prime(r)}{\rho_0(r)} \;,\label{eq:PDE2j}
\end{eqnarray}
where $\rho_0^\prime(r)=d\rho_0(r)/dr$. To solve Eq.~(\ref{eq:PDE2j}), two
boundary conditions are needed. The boundary condition at a distance
far away from the trap center can be obtained by noting that
\begin{eqnarray}
 \lim_{r\to\infty}\frac{\rho_0^\prime(r)}{\rho_0(r)} & = & \infty \;.\label{eq:denRatio}
\end{eqnarray}
Thus, the asymptotic value of the first derivative of the solution is
fixed by
\begin{eqnarray}
  \frac{d}{dr}\left[\frac{\tilde\phi_j(r)}{r^j}\right] \sim \frac{1}{r^{j+1}} & \rm for & r\to\infty \;.\label{eq:outBC}
\end{eqnarray}
The other boundary condition can be chosen as $\tilde\phi_j(0)=0$ because
the correction near the origin is small in a large system. With these
two boundary conditions, Eq.~(\ref{eq:PDE2j}) is solved numerically
to obtain $\tilde\phi_j(r)$, and thus the GP basis states.

The second step is to find the overlap between two GP basis states
because they are not orthogonal to each other. However, it is
sufficient to calculate the overlap between two single-particle wave
functions up to the second order in the separation between their
vortex coordinates because the corresponding many-body wave functions
are nearly orthogonal. The overlap between two single-particle wave
functions is
\begin{eqnarray}
\lefteqn{ \bra\GPw(\vr_0^\prime)|\GPw(\vr_0)\ket } \nonumber\\
& \approx & \int d^2\vr e^{i[\phi(\vr;\vr_0)-\phi(\vr;\vr_0^\prime)]}g(|\vr-\vr_0^\prime|)g(|\vr-\vr_0|)|\gsw(r)|^2 \nonumber\\
& \approx & \int d^2\vr e^{i[\vartheta(\vr;\vr_0)-\vartheta(\vr;\vr_0^\prime)]}g(|\vr-\vr_0^\prime|)g(|\vr-\vr_0|)|\gsw(r)|^2 \nonumber\\
&& - |\vr_0-\vr_0^\prime|^2\int \frac{d^2\vr}{4r^2}\left[2\tilde\phi_1(r)+\tilde\phi_1(r)^2\right]|\GPw(\vr;{\bf 0})|^2 \nonumber\\
& \approx &  1-\alpha_R\frac{|\vr_0-\vr_0^\prime|^2}{N\sigma^2}-i\alpha_I\frac{\hat{\vz}\cdot\vr_0\times\vr_0^\prime}{N\sigma^2} \;,\label{eq:ovl}
\end{eqnarray}
where $\alpha_R$ and $\alpha_I$ are numerical constants for a given system.  The
coefficient $\alpha_R$ generally scales logarithmically with the system
size, while the coefficient $\alpha_I$ is not sensitive to the system size
or local density variations, and is exactly given by $\pi$ in an
asymptotically uniform system~\cite{Tang2001b}.

Finally, the density profile of the state $|\Psi_{n,l}\ket$ is given by
\begin{widetext}
\begin{eqnarray}
\rho_{n,l}(\vr) & = & N\int d^2\vr_0^\prime\int d^2\vr_0 F_{n,l}^*(\vr_0^\prime)F_{n,l}(\vr_0)\GPw^*(\vr;\vr_0^\prime)\GPw(\vr;\vr_0)\exp\left[-\alpha_R\frac{|\vr_0^\prime-\vr_0|^2}{\sigma^2}-i\alpha_I\frac{\hat\vz\cdot\vr_0\times\vr_0^\prime}{\sigma^2}\right] \;,
\end{eqnarray}
\end{widetext}
where the many-body overlap corresponding to the single-particle
overlap in Eq.~(\ref{eq:ovl}) falls off exponentially with the
separation of the vortex coordinates in the large $N$ limit.  To
compare this two-dimensional theory with experimental data, we assume
that the healing length $\xi$ and the interatomic spacing $\sigma$ at the
trap center is the same as in the three-dimensional system. In the
following, I will use the experimental conditions of
\citet{Madison2000a} (The trapping frequencies are listed in Table
\ref{tb:ratio}). For a BEC with $N_{\rm 3D}=1.4\times 10^5$ particles, the
healing length is $\xi=0.19$ $\mu$m and the interatomic spacing is
$\sigma=0.17$ $\mu$m. The corresponding number of particles in two dimensions
given by Eq.~(\ref{eq:Np2D}) is about $350$. The two coefficients in
the overlap are $\alpha_R=5.84$ and $\alpha_I=3.25$.

Although all states described by Eq.~(\ref{eq:Usol}) can exist, some
of them are more stable and likely to be realized in actual
experiments than the others.  A rotating BEC is typically created by a
stirring laser beam. The system is first driven by the laser stirrer
for a period of time, and then relaxes to a stationary state. The
laser stirrer breaks the rotational symmetry, and transfers both
energy and angular momentum into the system. During the relaxation
period, the energy can relax through collisional processes, but it is
harder for the angular momentum to relax unless the rotational
symmetry is broken by other means. There is experimental evidence
showing that the system can maintain its angular momentum for a long
period of time~\cite{Bretin2003a}.  This leads us to assume that the
system will fall into the lowest energy state with the initially given
angular momentum. As the angular momentum of the system increases,
more and more vortices will form. However, it is energetically
favorable to excite the existing vortices if the change in angular
momentum is small. The energy cost for introducing an additional
vortex into the system is approximately $\pi\ln(R/\xi)\hbar^2/m\sigma^2$, where $R$
is the characteristic radius of the system, while the energy cost for
having an excited vortex is approximately proportional to $n\hbar^2/m\sigma^2$,
where $n$ is the angular momentum increase and the proportionality
constant is less than one~\cite{Tang2001a,Tang2001b}. Although adding
another vortex would eventually become energetically favorable for
larger values of the angular momentum, we do not know what the
critical value is because accurate comparison of the energies of
different vortex configurations is complicated by the nonuniformity of
the system and require further studies.

In the case of a single excited vortex, the lowest energy state for a
given angular momentum is the $|\Psi_{n,n}\ket$ state. Here the density
profiles of the $|\Psi_{n,n}\ket$ states are compared against
experimental data.  Figure \ref{fig:denVIH}(a) shows that the data of
\citet{Madison2000a} can be best fitted with the state that the
angular momentum is increased by $5\%$ ($n=18$ compared to $N=350$).
The experimental data is shown by the solid dots. Only one centered
vortex core is clearly visible, and the core density profile is
significantly deviated from the density profile of the GP state
($L_z=N\hbar$).  Fig.~\ref{fig:denVIH}(b) shows the trend that the core
density increases with increasing angular momentum of the system
($L_z=(N+n)\hbar$).

In summary, I have calculated the density profiles of different
rotational states of a trapped Bose gas with one quantized vortex
using many-body wave functions which are linear combinations of GP
wave functions. The core density increases with increasing angular
momentum of the system, which suggests that vortices with large core
densities are in excited states with higher angular momentum. The
possibility of having excited vortices suggests that for given angular
momentum, the lower energy rotational state can have number of
vortices less than expected in the mean-field theory. From the
quantitative agreement on the density profile between the theoretical
prediction and the experimental measurements, it is suggested that the
rotating Bose gas observed experimentally by \citet{Madison2000a} is
in a state with angular momentum higher than one $\hbar$ per atom.

The author thanks D. J. Thouless for valuable discussions, and K. W.
Madison and J. Dalibard for providing data shown in
Fig.~\ref{fig:denVIH}.

\end{document}